\documentclass[aps,prl,amsmath,amssymb,preprint]{revtex4-1}
\pdfoutput=1
\usepackage{graphicx}
\usepackage{dcolumn}
\usepackage{bm}
\usepackage{amsmath}
\usepackage[dvipsnames]{xcolor}
\usepackage{multirow}
\usepackage{changes}
\usepackage{endnotes}
\usepackage{textgreek}
\usepackage{cases}
\usepackage[T1]{fontenc}
\usepackage[utf8]{inputenc}

\begin{document}

\title{Ultrafast dynamics in (TaSe$_4$)$_2$I triggered by valence and
core-level excitation}

\author{Wibke Bronsch}
\email[Corresponding author: ]{wibke.bronsch@elettra.eu}
\affiliation{Elettra - Sincrotrone Trieste S.C.p.A., Strada Statale 14 - km 163.5 in AREA Science Park, 34149 Basovizza, Trieste, Italy}

\author{Manuel Tuniz}
\affiliation{Dipartimento di Fisica, Universit\`{a} degli Studi di Trieste, 34127 Trieste, Italy}

\author{Giuseppe Crupi}
\affiliation{Dipartimento di Fisica, Universit\`{a} degli Studi di Trieste, 34127 Trieste, Italy}

\author{Michela De Col}
\affiliation{Dipartimento di Fisica, Universit\`{a} degli Studi di Trieste, 34127 Trieste, Italy}

\author{Denny Puntel}
\affiliation{Dipartimento di Fisica, Universit\`{a} degli Studi di Trieste, 34127 Trieste, Italy}

\author{Davide Soranzio}
\affiliation{Dipartimento di Fisica, Universit\`{a} degli Studi di Trieste, 34127 Trieste, Italy}

\author{Alessandro Giammarino}
\affiliation{Dipartimento di Fisica, Universit\`{a} degli Studi di Trieste, 34127 Trieste, Italy}

\author{Michele Perlangeli}
\affiliation{Dipartimento di Fisica, Universit\`{a} degli Studi di Trieste, 34127 Trieste, Italy}

\author{Helmuth Berger}
\affiliation{Institute of Physics, Ecole Polytechnique F\'{e}d\'{e}rale de Lausanne (EPFL), CH-1015 Lausanne, Switzerland}

\author{Dario De Angelis}
\affiliation{Elettra - Sincrotrone Trieste S.C.p.A., Strada Statale 14 - km 163.5 in AREA Science Park, 34149 Basovizza, Trieste, Italy}

\author{Danny Fainozzi}
\affiliation{Elettra - Sincrotrone Trieste S.C.p.A., Strada Statale 14 - km 163.5 in AREA Science Park, 34149 Basovizza, Trieste, Italy}

\author{Ettore Paltanin}
\affiliation{Elettra - Sincrotrone Trieste S.C.p.A., Strada Statale 14 - km 163.5 in AREA Science Park, 34149 Basovizza, Trieste, Italy}
\affiliation{Dipartimento di Fisica, Universit\`{a} degli Studi di Trieste, 34127 Trieste, Italy}

\author{Stefano Pelli Cresi}
\affiliation{Elettra - Sincrotrone Trieste S.C.p.A., Strada Statale 14 - km 163.5 in AREA Science Park, 34149 Basovizza, Trieste, Italy}

\author{Gabor Kurdi}
\affiliation{Elettra - Sincrotrone Trieste S.C.p.A., Strada Statale 14 - km 163.5 in AREA Science Park, 34149 Basovizza, Trieste, Italy}

\author{Laura Foglia}
\affiliation{Elettra - Sincrotrone Trieste S.C.p.A., Strada Statale 14 - km 163.5 in AREA Science Park, 34149 Basovizza, Trieste, Italy}
 
\author{Riccardo Mincigrucci}
\affiliation{Elettra - Sincrotrone Trieste S.C.p.A., Strada Statale 14 - km 163.5 in AREA Science Park, 34149 Basovizza, Trieste, Italy}

\author{Fulvio Parmigiani}
\affiliation{Elettra - Sincrotrone Trieste S.C.p.A., Strada Statale 14 - km 163.5 in AREA Science Park, 34149 Basovizza, Trieste, Italy}

\author{Filippo Bencivenga}
\affiliation{Elettra - Sincrotrone Trieste S.C.p.A., Strada Statale 14 - km 163.5 in AREA Science Park, 34149 Basovizza, Trieste, Italy}

\author{Federico Cilento}
\email[Corresponding author: ]{federico.cilento@elettra.eu}
\affiliation{Elettra - Sincrotrone Trieste S.C.p.A., Strada Statale 14 - km 163.5 in AREA Science Park, 34149 Basovizza, Trieste, Italy}

\begin{abstract}
Dimensionality plays a key role for the emergence of ordered phases such as charge-density-waves (CDW), which can couple to, and modulate, the topological properties of matter.
In this work, we study the out-of-equilibrium dynamics of the paradigmatic quasi-one-dimensional material (TaSe$_4$)$_2$I, that exhibits a transition into an incommensurate CDW phase when cooled just below room temperature, namely at T$_{\rm{CDW}} $= 263\,K. 
We make use of both optical laser and free-electron laser (FEL) based time-resolved spectroscopies in order to study the effect of a selective excitation on the normal-state and on the CDW phases, by probing the near-infrared/visible optical properties both along and perpendicularly to the direction of the CDW, where the system is metallic and insulating, respectively. 
Excitation of the core-levels by ultrashort X-ray FEL pulses at 47 eV and 119 eV induces reflectivity transients resembling those recorded when only exciting the valence band of the compound - by near-infrared pulses at 1.55 eV - in the case of the insulating sub-system. 
Conversely, the metallic sub-system displays a relaxation dynamics which depends on the energy of photo-excitation. 
Moreover, excitation of the CDW amplitude mode is recorded only for excitation at low-photon-energy. 
This fact suggests that the coupling of light to ordered states of matter can predominantly be achieved when directly injecting delocalized carriers in the valence band, rather than localized excitations in the core levels. 
On a complementary side, table-top experiments allow us to prove the quasi-unidirectional nature of the CDW phase in (TaSe$_4$)$_2$I, whose fingerprints are detected along its $c$-axis only. 
Our results provide new insights on the symmetry of the ordered phase of (TaSe$_4$)$_2$I perturbed by a selective excitation, and suggest a novel approach based on complementary table-top and FEL spectroscopies for the study of complex materials.
\end{abstract}

\maketitle
\date{\today}

\section{Introduction}
\begin{figure}[t]
	\centering
	\includegraphics[width=0.5\textwidth]{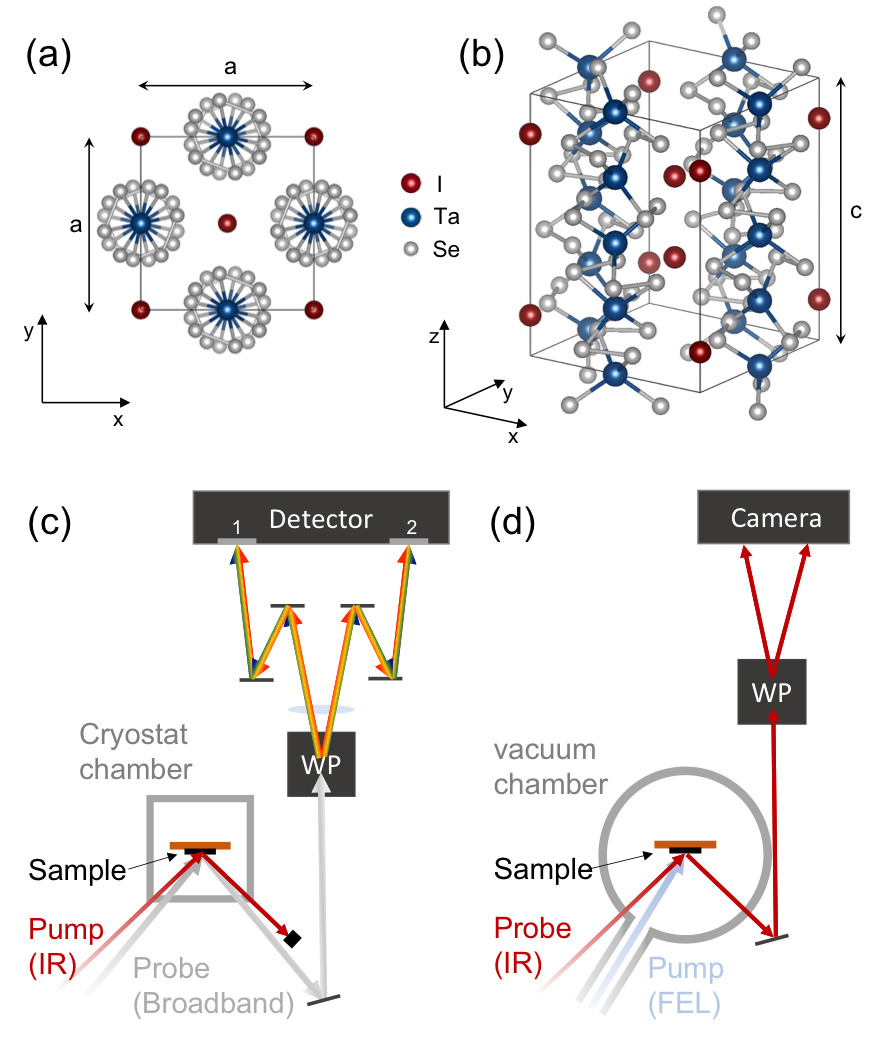}
	\caption{\footnotesize Crystal structure and experimental geometries. 
	a) Top view of the unit cell of (TaSe$_4$)$_2$I. 
	b) Unit cell of (TaSe$_4$)$_2$I; the in-plane axis, $a$, and the $c$ axis (direction of the quasi-1D CDW) are indicated. The crystal structure displayed in a) and b) has been produced by using VESTA. 
	All the lattice parameters used to perform these plots are freely accessible at https://materialsproject.org/materials/mp-30531/. 
	c) Setup for table-top optical pump-probe spectroscopy with supercontinuum probe. d) Setup for pumping with FEL light and probing at $\lambda$=800\,nm. With both setups, the $a$-axis and $c$-axis time-resolved optical properties are measured simultaneously. 
	Note that in panel c) and d) the beams are depicted at arbitrary incident angles for simplifying the sketches. In both setups the probe beam is impinging on the sample in near-normal incidence geometry.}
	\label{fig:setups}
\end{figure}

Non-equilibrium spectroscopy techniques developed in a short period of time from avant-garde experiments to an established tool for studying the properties of complex materials, delivering comprehensive datasets, combining both spectroscopic and temporal information. 
This allows for revealing and disentangling electron, lattice and spin dynamics, facilitating the comprehension of cause-effect relationships and the interplay among the various degrees of freedom \cite{Giannetti2016}.
A common ansatz to gain such information is perturbing an ordered state by an intense, ultrashort light pulse, below the sample damage threshold, and tracing the material response during its relaxation back to equilibrium conditions.
Especially in connection to materials displaying phase transitions, non-equilibrium spectroscopies received great interest in order to gain insights about the driving forces responsible of the transition. 
Examples comprise the study of correlated materials and superconductors \cite{Giannetti2011}, charge-density-wave (CDW) systems \cite{Hellmann2012,Payne2020} and exciton-related dynamics \cite{Mor2017,Mor2021}. 
In several systems, photo-induced phase-transitions \cite{Wegkamp2015} can switch the physical and functional properties of the material on ultrafast timescales.
Optical pump-probe spectroscopy, which allows for studying effects on the valence band structure, deliver comprehensive information about various effects as {\it e.g.} band gap renormalization \cite{Jnawali2020}, the excitation of coherent phonons \cite{Soranzio2019} or the excitation of the amplitude mode in CDW materials \cite{Mann2016}.
Depending on the ground state of the compound under scrutiny, and the intensity (and photon energy) of the photoexcitation, the system can evolve in a new state by following different pathways, that may or may not have a counterpart at equilibrium \cite{Kogar2020,Zhou2021}.\\
X-ray pulses, in contrast to light pulses in the visible range, open the possibility to study many-body effects triggered by the creation of a core hole. 
Ultrafast changes of the optical reflectivity induced by X-ray pulses aroused great interest as a tool for femtosecond X-ray/optical cross-correlation, which serves as a beam characterization tool for more widely spread IR-pump/X-ray probe experiments.\cite{Gahl2008,Eckert2015}
Less explored than the use of X-ray probes in order to make use of the element selectivity of synchrotron and free-electron laser radiation, is the ability of intense X-ray pulses to alter the refractive index of solids on an ultrafast timescale. 
Because of the absorption of x-rays, in addition to the excitation of the valence band, photoelectrons and core holes are generated.
The presence of core level holes induces Auger or fluorescence processes, which ultimately alter the carrier density in the conduction and valence bands and thus perturbing the optical reflectivity.\cite{Durbin2012}
Even though optical reflectivity changes induced by intense X-ray pulses offer the possibility to study ultrafast many body responses, so far the variety of materials studied by applying this technique is rather small and the interpretation and analysis of the experimental data is a recently discussed topic.\cite{Durbin2012,Casolari2014,Eckert2015,Tkachenko2016}
\\
In the present work, we report on a comprehensive study on the effect of core-level and valence band excitations on the paradigmatic quasi-one-dimensional material \cite{Monceau2012} (TaSe$_4$)$_2$I, tracing the changes in its optical reflectivity induced by photoexcitation with ultrashort light pulses. In particular, we compare the influence of core-level and valence band excitations on the optical properties of (TaSe$_4$)$_2$I single crystals along both their $a$- and $c$-axes. By doing so, we can study the effect of photoexcitation at considerably different photon energy simultaneously on a metallic and an insulating band structure.
\\
(TaSe$_4$)$_2$I has recently gained a significant attention due to its rich topology.\cite{Shi2021,Yi2021,Li2021}
To our aim, (TaSe$_4$)$_2$I constitutes an ideal model system because it allows us to compare the influence of valence band and core-level excitations on different electronic structures. 
Indeed, as anticipated above, in the normal state the material obeys a metallic band structure along its $c$-axis and insulating properties along the $a$-axes.\cite{Shi2021} 
Figures\,\ref{fig:setups}a) and b) show a top and a side view of the unit cell of the crystal structure.
Below a critical temperature of T$_{\rm C}$=263\,K, the material undergoes a phase transition into a CDW phase.\cite{Wang1983,Gooth2019}
Due to the quasi one-dimensional character of the phase-transition, the periodic lattice distortion related to the CDW phase sets-in along the $c$-axis of the crystal and leaves the $a$-axes non-impacted. Schaefer et al. \cite{Schaefer2013} first detected fingerprints of the phase transition from the metallic to the CDW phase of (TaSe$_4$)$_2$I in the time-resolved optical properties, when the material is excited in the near-infrared range (at 800\,nm or at 1300\,nm).
On the contrary, the effect of core-hole excitations on the valence and conduction band structure of (TaSe$_4$)$_2$I has not been investigated yet.
Here we show for the first time the direct comparison of the non-equilibrium dynamics along the different crystal axis and for different excitation schemes, fluences and temperatures.

\section{Experimental Setups}
We studied the optical response of (TaSe$_4$)$_2$I to valence bands and core-level excitations by using two different setups, which are schematically shown in Figs.\,\ref{fig:setups}c),d). 
Figure\,\ref{fig:setups}c) illustrates the table-top optical pump-probe spectroscopy setup available at the T-ReX facility at Elettra - Sincrotrone Trieste.
We worked with a pulsed Ti-Sapphire source at a repetition rate of 250\,kHz. 
As a pump wavelength we used the fundamental of the source centered at $\approx$800\,nm, whereas for the probe beam we used a supercontinuum beam to be able to probe changes in the reflectivity of the sample in the spectral range $\approx$450-900\,nm. 
The broadband probe beam hit the sample at near-normal incidence, while the pump beam impinged on the sample with an angle of $\approx$20$^{\circ}$ to the surface normal. 
We measured the spot-size to be around 40\,$\mu$m FWHM (Full Width at Half Maximum) for the probe beam and to $\approx$140\,$\mu$m FWHM for the pump beam.
The reflected beam from the sample was lead through a Wollaston prism, which allowed for simultaneous detection of the sample response along and perpendicular to the direction of the CDW.
Details about the setup and the supercontinuum source are published in Ref.\cite{Perlangeli2020}.
During the measurements we kept the sample mounted in a high vacuum chamber. 
The samples were glued to a copper stick attached to a cryostat, which allowed for measurements at variable temperatures down to 15\,K in the case of liquid helium cooling.

Figure\,\ref{fig:setups}d) schematically shows the setup used for core-level excitations.
Free electron laser (FEL) pump and 800\,nm probe experiments were performed at the FEL FERMI in Trieste.\cite{Bencivenga2015}
Due to the opacity of the investigated samples to IR light, all experiments were performed in reflection geometry. 
The IR-probe beam was impinging on the sample in a near-normal incidence geometry, whereas the FEL pump beam arrived with an angle of $\approx$40$^{\circ}$ with respect to the surface normal in the case of a photon energy of 47\,eV and of 13.5$^{\circ}$ in the case of 119\,eV photon energy, respectively.
The reflected IR light from the sample was split by a Wollaston prism in two components, which relate to the reflectance along and perpendicular to the CDW chains.
In this way we were able to simultaneously record the optical response of the sample to the FEL excitation along the two high-symmetry axes of (TaSe$_4$)$_2$I.
The FEL beam had a spot-size of 150\,\textmu m x 200\,\textmu m FWHM, whereas the probing 800\,nm had a FWHM of 150\,\textmu m x 150\,\textmu m.
The repetition rate of the FEL amounted to 50\,Hz.
The needle-like (TaSe$_4$)$_2$I crystalline samples were synthesized by the CVD (chemical vapor transport) technique starting from a stoichiometric mixture of high-purity elements. Their size was of the order $6.0\times 0.3 \times 0.3$ mm$^3$.
\newpage


\begin{figure*}
\centering
\includegraphics[width=0.92\textwidth]{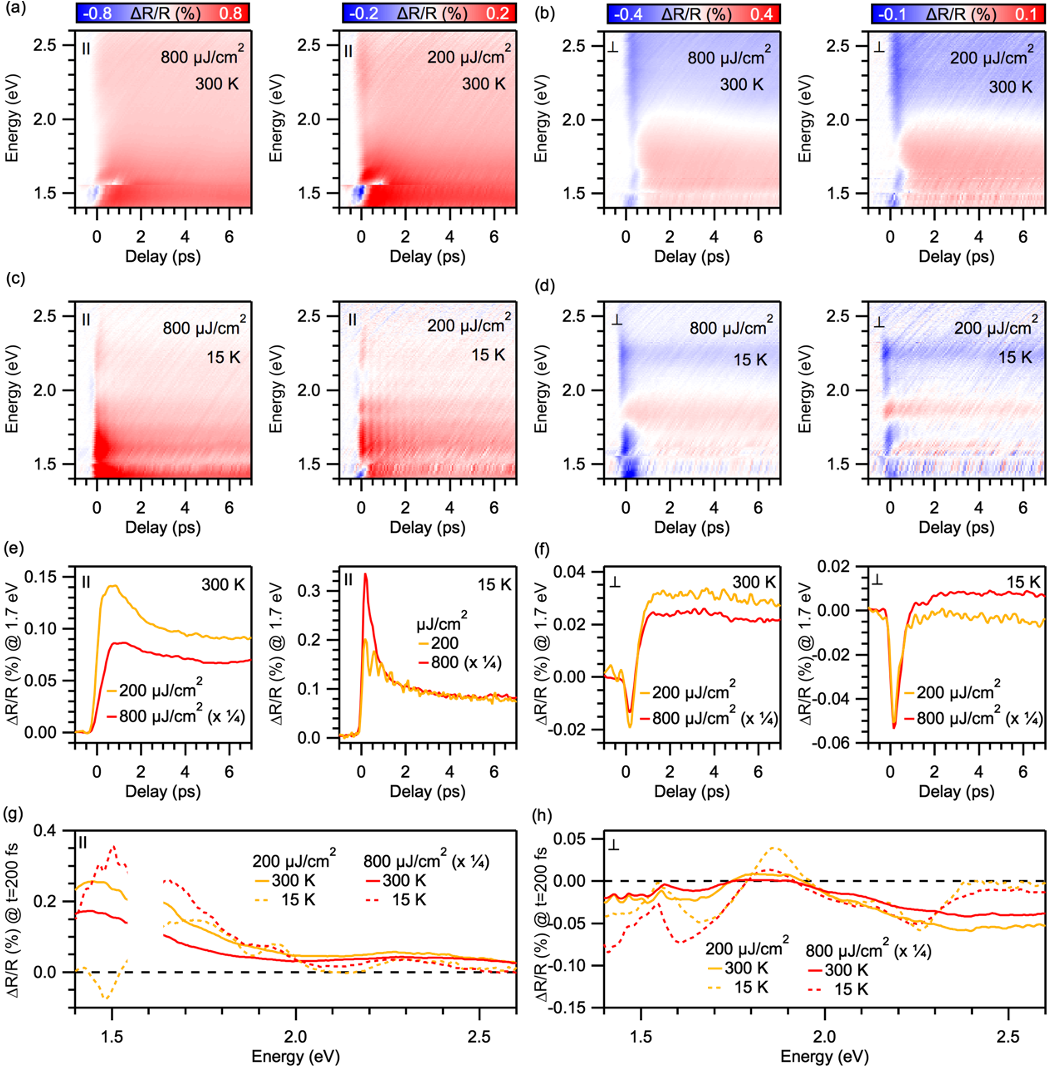}
\caption{\footnotesize Time-Resolved Optical Spectroscopy of (TaSe$_4$)$_2$I. 
Time-and-frequency resolved normalized reflectivity variations at T=300\,K, a) for the parallel direction ($c$-axis) and b) for the perpendicular direction ($a$-axis). 
Panels c) and d) refer to the case T=15\,K. 
All panels comprise a measurement acquired at an absorbed fluence of 800\,$\mu$J/cm$^2$ and at an absorbed fluence of 200\,$\mu$J/cm$^2$. 
The extrema of the colorscale of the measurements acquired at low fluence are set to $1/4$ of the ones of the measurements at high fluence for ease comparison. 
All the measurements acquired at the same fluence and along the same direction share the same colorscale. 
The colorscales for the perpendicular direction range half of the ones for the parallel direction for better visibility of the changes.
e) Time cuts extracted at $h\nu$=1.7\,eV ($\lambda$=730\,nm) for the parallel direction, at T=300\,K and T=15\,K, for the two fluence values considered. 
f) Same as e) but for the perpendicular direction. 
g) Spectral profiles extracted at t=200\,fs for the parallel direction, at T=300\,K and T=15\,K, for the two fluence values considered. 
g) Same as f) but for the perpendicular direction.}
\label{fig:TROS_data}
\end{figure*}
\newpage
\section{Results and Discussion}
\subsection{Optical Excitation}
\label{sec:optical pump-probe}
We performed time-resolved optical spectroscopy experiments on high-quality (TaSe$_4$)$_2$I crystals, in order to study the effect of photoexcitation at $h\nu_P$=1.55\,eV on the electronic band structure of the compound, for different conditions of temperature (above and below the critical temperature T$_c\sim$263\,K) and fluence. 
Following the sketches reported in Fig.\,\ref{fig:setups}, and carefully aligning the sample, the setup allowed us to acquire simultaneously the transient reflectivity variation, $\Delta R/R$, for both the $c$-axis component (z direction, termed in the following "parallel"-direction, that is, the direction parallel to the quasi-1D CDW) and the $a$-axis component (x or y direction, termed "perpendicular" direction). 
Spectroscopic measurements are shown in Fig.\,\ref{fig:TROS_data}a)-d) as bi-dimensional maps where the $\Delta R/R$, indicated as a colorscale, is plotted versus pump-probe delay t and probe photon energy $h\nu$, that is: $\Delta R/R$(t,$h\nu$). 
Panels a) and b) show the data acquired above T$_c$, at T=300\,K, for the parallel and perpendicular component respectively. 
Each panel comprises two measurements acquired at different (absorbed) fluences, 800\,$\mu$J/cm$^2$ and 200\,$\mu$J/cm$^2$. 
The extrema of the colorscale for the low-fluence measurements are set to $1/4$ of the ones of the high-fluence measurements, in order to facilitate comparisons. 
Similarly, panels c) and d) display the data acquired well below T$_c$, at T=15\,K, for the parallel and perpendicular component respectively. 
The parallel, metallic-like component and the perpendicular, insulating-like component, show markedly different behaviors, that will be addressed by a dedicated analysis described in the following. 
Briefly, the parallel component displays a positive signal which peaks in the near-infrared side of the measured spectral range and decreases smoothly towards the visible range. 
Conversely, the perpendicular component shows a spectrally-structured $\Delta R/R$ signal with sign changes, which is smaller in amplitude than the parallel-one. 
Upon cooling, the spectral features become sharper, and coherent oscillations linked to the excitation of the amplitude mode of the CDW are detected along the parallel direction at low fluence.
We want to point out here that the $\Delta R/R$(t,$h\nu$) maps displayed in panels a)-d) have been corrected for the supercontinuum probe pulse chirp. The correction consists in the alignment of the arrival time of each spectral component of the pulse. 
However, in the case of the parallel component, it is clear from the data of panels a)-d) that a proper chirp correction in the vicinity of $h\nu$=1.55 eV ($\lambda$=800 nm) cannot be achieved. 
For this reason, for the parallel component data only, the energy region 1.5-1.65 eV will be excluded from the analysis.
For a qualitative analysis of the response of (TaSe$_4$)$_2$I to optical photoexcitation, time profiles (at constant $h\nu$) and spectral profiles (for a given delay t) are extracted from the $\Delta R/R$(t,$h\nu$) maps. 
In panels e) and f) we are reporting time profiles exemplary taken at $h\nu$=1.7\,eV ($\sim$730\,nm). 
The traces are scaled by the fluence's ratio. 
We observe that at T=300\,K both components show a sub-linear dependence on the fluence.
Conversely, at T=15\,K, the trend is opposite, and the signal is super-linear. 
For the metallic, parallel component, we observe a double-exponential decay (faster at low temperature), while for the insulating perpendicular component a positive plateau is reached after a fast, negative transient. 
The most important fluence-dependent effect is observed for the parallel component at T=15\,K, where the coherent oscillations are almost completely washed out after photoexcitation at 800\,$\mu$J/cm$^2$, signaling that the CDW phase can be melted by impulsive photoexcitation. 
The spectral profiles extracted at t$=200$\,fs are reported in panels g) and h), and confirm that the spectral features in the perpendicular component become sharper at lower temperature.
\begin{figure}[t]
\centering
\includegraphics{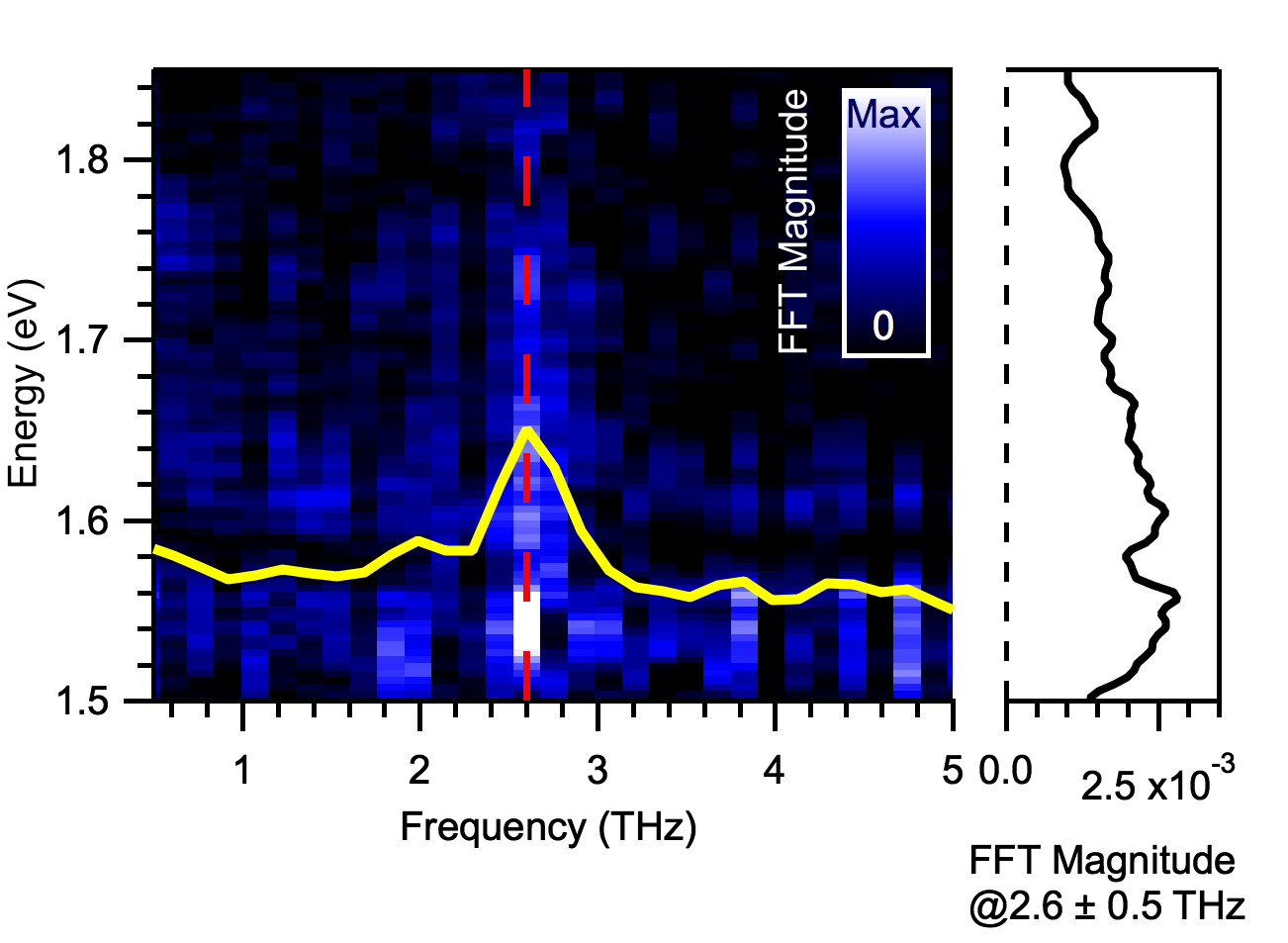}
\caption{\footnotesize I($\omega$,$h\nu$) map displaying the FFT of the coherent component extracted from the $\Delta R/R$(t,$h\nu$) map acquired at T=15\,K and 200\,$\mu$J/cm$^2$ fluence. 
The photon energy range 1.5-1.85\,eV is considered, where the coherent oscillations associated to the excitation of the amplitude mode are more pronounced. 
The yellow curve, obtained by integrating the FFT amplitude over the full photon energy range considered, indicates the oscillation frequency. 
It peaks at 2.6\,THz, as marked by the red dashed line. The black curve shows the photon-energy-dependent FFT amplitude, as integrated in a 0.5\,THz-wide interval centered at 2.6\,THz.}
\label{fig:FFT}
\end{figure}
The results of the analysis of the amplitude mode, as obtained from the dataset reported in Fig.\,\ref{fig:TROS_data}c) at 200\,$\mu$J/cm$^2$ fluence, are reported in Fig.\,\ref{fig:FFT}.
In particular, the map I($\omega$,$h\nu$) shows the spectral evolution of the FFT (Fast Fourier Transform) of the residuals (coherent component only) of the $\Delta R/R$(t,$h\nu$) map. 
The residuals have been computed by subtracting to each time profile at constant photon energy a fit of the same trace obtained with a double-exponential decay function (convoluted with a Gaussian for accounting the finite time resolution), that represents the incoherent response. 
By stacking the FFT of the residuals we obtain the map reported in Fig.\,\ref{fig:FFT}, that displays the evolution of the oscillation frequency as a function of the probe photon energy. A peak at 2.6\,THz is clearly detected, which corresponds to the amplitude mode of the CDW in (TaSe$_4$)$_2$I. 
This finding is in agreement with earlier results reported in \cite{Schaefer2013}, for $\lambda$=800\,nm. 
In our case, we gain information about the spectral dependence of the coherent modulation, which extends in a wide spectral range from the near-infrared beyond $\approx$2\,eV (cf. data shown in Fig.\,\ref{fig:TROS_data}c); 200\,$\mu$J/cm$^2$).

\begin{figure*}[t]
\centering
\includegraphics[width=0.92\textwidth]{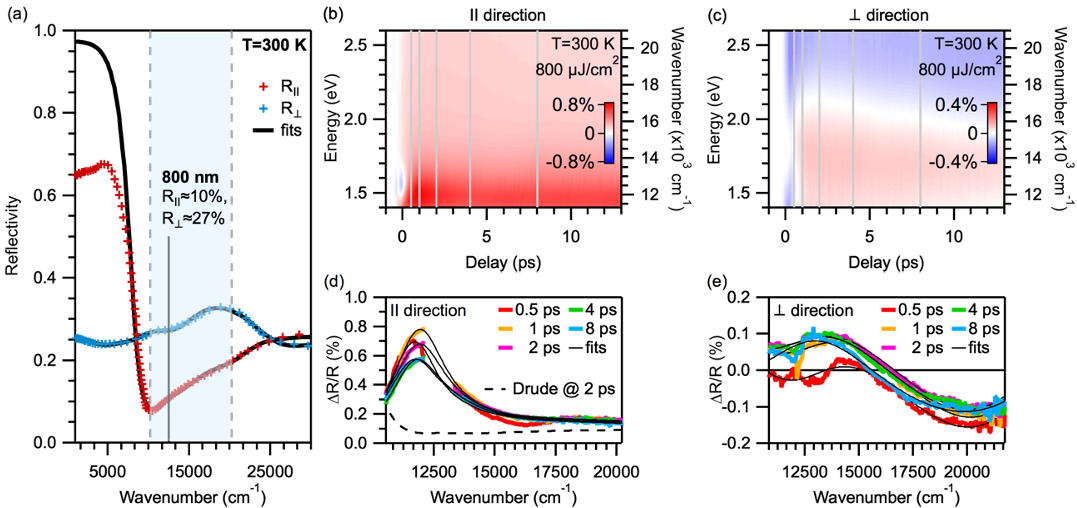}
\caption{\footnotesize Analysis of the time-resolved optical spectroscopic data acquired at T=300\,K. 
a) Reflectivity along the crystal $a$- and $c$-axis at equilibrium, reported by Berner {\it et al.}\cite{Berner1993}. The spectral range probed by the setup sketched in Fig.\,\ref{fig:setups}c is emphasized. 
The photon energy $h\nu$=1.55\,eV (12500 cm$^{-1}$), corresponding to $\lambda$=800\,nm, is marked. 
This corresponds to the pump photon energy of the setup in Fig.\,\ref{fig:setups}c and the probe photon energy of the setup in Fig.\,\ref{fig:setups}d. 
The fits to the two reflectivity curves have been obtained as described in the main text.
Panel b) displays the result of the fit of a Drude-Lorentz model to the data acquired along the parallel direction. 
The map is obtained by stacking the fits performed at each pump-probe delay. 
Panel c) displays the same quantity, but for the perpendicular component. 
The details of the analysis are described in the main text. 
The vertical gray lines shown in panels b) and c) mark a few selected time delays at which the fit profiles are extracted and compared to the corresponding spectral profile of the $\Delta R/R$(t,$h\nu$) map. 
The fit profiles (displayed as solid black lines) are reported in panels d) and e) for the parallel and perpendicular component respectively, while the spectral profiles are displayed with colored curves indicating the time delay. 
An overall good agreement between data and model is achieved both in the vicinity of t$_0$ and at large time delays. 
For the spectral profiles of the parallel direction, the energy interval (1.5-1.65\,eV) has been excluded from the fit because of the difficulty in correcting the probe pulse chirp in the vicinity of $h\nu$=1.55\,eV.}
\label{fig:TROS_fit}
\end{figure*}

\begin{figure*}[t]
\centering
\includegraphics[width=0.92\textwidth]{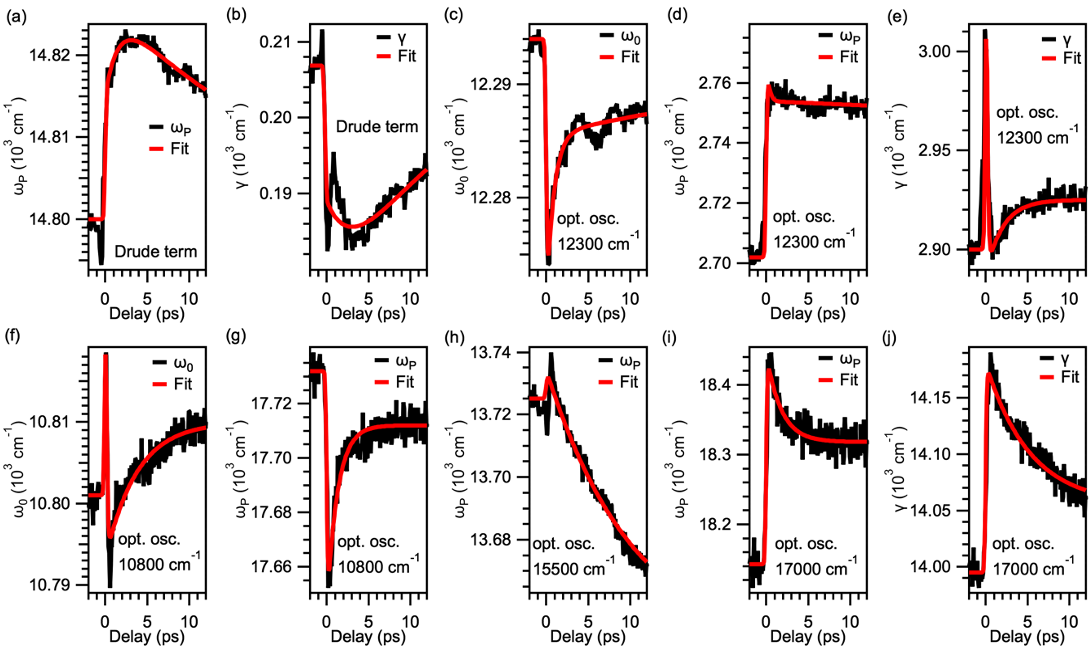}
\caption{\footnotesize Dynamics of the parameters of the Dielectric Function, as obtained from the fitting procedure described in the main text. 
Panels a)-e) refer to the parameters of the dielectric function for the parallel-direction optical properties. 
Panels f)-j) refer instead to the perpendicular-direction optical properties. 
a) Plasma frequency of the Drude term. 
b) Scattering Rate of the Drude term. 
c) Central wave-number of the optical oscillator at 12300\,cm$^{-1}$. 
d) Plasma frequency of the optical oscillator at 12300\,cm$^{-1}$. 
e) Linewidth of the optical oscillator at 12300\,cm$^{-1}$. 
f) Central wave-number of the optical oscillator at 10800\,cm$^{-1}$. 
g) Plasma  Frequency of the optical oscillator at 10800\,cm$^{-1}$. 
h) Plasma  Frequency of the optical oscillator at 15500\,cm$^{-1}$. 
i) Plasma  Frequency of the optical oscillator at 17000\,cm$^{-1}$. 
j) Linewidth of the optical oscillator at 17000\,cm$^{-1}$.}
\label{fig:parameters}
\end{figure*}

In order to gain quantitative insight about the microscopic origin of the $\Delta R/R(h\nu$) profiles, we developed a model for the dielectric function of (TaSe$_4$)$_2$I and used a differential approach to find which of its components are modified by photoexcitation. 
The prerequisite for this analysis is the detailed comprehension of the optical properties at equilibrium. 
The reflectivity of (TaSe$_4$)$_2$I at T=300\,K was measured by Berner {\it et al.}\cite{Berner1993}. 
We digitized the reflectivity curves for both parallel and perpendicular components, which are reported in Fig.\,\ref{fig:TROS_fit}a). 
The solid lines are the fit to the reflectivity curves of a Drude-Lorentz model for the dielectric function.
The reflectivity at near-normal incidence is computed using the expression R=$|(1-\sqrt\epsilon)/(1+\sqrt\epsilon)|^2$, where $\epsilon$=$\epsilon$($h\nu$) is the model complex dielectric function. The expression for $\epsilon$($h\nu$) in terms of Drude and Lorentz optical oscillators is extensively discussed and reported in \cite{Wooten}.
The reflectivity curve for the direction parallel to the CDW clearly shows the presence of a plasma edge, which is instead absent in the perpendicular component. 
Hence, for the dielectric function model we used a Drude term only for the parallel component. 
Additionally, a number of Lorentz oscillators was added in order to obtain a satisfactory fit of the reflectivity curves, using the smallest number of oscillators for minimizing the $\chi^2$ of the fit. 
For the parallel component, four Lorentz oscillators are required; they have central wave-numbers of 12300\,cm$^{-1}$, 17000\,cm$^{-1}$, 22050\,cm$^{-1}$, 27880\,cm$^{-1}$, respectively. 
For the perpendicular component, a total of six Lorentz oscillators is required, having central wave-numbers of 4500\,cm$^{-1}$, 7000\,cm$^{-1}$, 10800\,cm$^{-1}$, 15500\,cm$^{-1}$, 17000\,cm$^{-1}$, 28200\,cm$^{-1}$, respectively.
Once the optical properties at equilibrium were described by a model dielectric function (one for each direction), the time-resolved optical properties were analyzed by computing the $\Delta R/R$($h\nu$) as $\Delta R/R(h\nu)=(R_{\rm neq}(h\nu)-R_{\rm eq}(h\nu))/R_{\rm eq}(h\nu)$, where $R_{\rm eq}(h\nu)$ was calculated from the model dielectric function described above, and $R_{\rm neq}(h\nu)$ was calculated from a 'copy' of the equilibrium model in which a few parameters were allowed to vary when the quantity $\Delta R/R$($h\nu$) was fitted to the spectral profiles.
We applied this analysis to the results obtained at T=300\,K, for consistence with the experimental conditions of the equilibrium spectroscopy, for excitation at 800\,$\mu$J/cm$^2$ fluence. In the analysis, the parallel and perpendicular dielectric functions were considered as independent.
The results are reported in Fig.\,\ref{fig:TROS_fit}. 
Panels b) and c) show the results of the fitting procedure for the parallel and perpendicular component respectively. 
They should be compared to the corresponding datasets reported in panels a) and b) of Fig.\,\ref{fig:TROS_data}. 
In order to discuss about the agreement of the fit to the data, a few profiles are extracted at selected fixed delays, marked by the vertical dashed lines superimposed to the maps. 
They are reported in panels d) and e) for the parallel and perpendicular component respectively, together with the corresponding spectral profile extracted from the datasets reported in Fig.\,\ref{fig:TROS_data} at the same time delays. 
The agreement is satisfactory both in the vicinity of t$_0$ and at large time delays, for both components. The fittings have been obtained as follows: 
For the parallel component, the Drude term (plasma frequency and scattering rate) and the oscillator at 12300\,cm$^{-1}$ (plasma frequency, central frequency and linewidth) were set as free parameters. 
Their temporal evolution and associated dynamics are shown in panels a)-e) of Fig.\,\ref{fig:parameters}. 
For the perpendicular component, the analysis reveals that three Lorentz oscillators are affected by photoexcitation, in particular those at 10800\,cm$^{-1}$, 15500\,cm$^{-1}$, 17000\,cm$^{-1}$. 
The number of parameters that has to be modified for a satisfactory fit is five, and in particular they are the central frequency and the plasma frequency of the first, the plasma frequency of the second, and the plasma frequency and linewidth of the third. 
Their time evolution is displayed in panels f)-j) of Fig.\,\ref{fig:parameters}. 
Before approaching the discussion of the dynamics, it is worth noting that some parameters show a non-trivial time-evolution in the vicinity of time-zero, that we associate to the little imperfections in the correction of the supercontinuum pulse chirp. 
For this reason, we will focus mainly on the discussion of the dynamics on the picosecond timescale.
The knowledge of which parameter of the dielectric function is altered by photoexcitation, and how, allows us to understand the effect of photoexcitation at the microscopic level. 
For the parallel component, photoexcitation affects the free carriers in the conduction band, which are described by the Drude term. 
In particular, the plasma frequency increases, signaling that extra carriers are injected in the conduction band. 
This is accompanied by a reduction of the scattering rate. 
This is a counter-intuitive effect (heating typically broadens the plasma edge, i.e., the scattering rate increases) that asks for more detailed investigations in order to determine the reason behind it. 
The dielectric function analysis allows us to isolate the effect of the modification of the Drude term alone on the $\Delta R/R$ signal. 
We did this for t=2\,ps, by picking up the values for the Drude term in the non-equilibrium dielectric function, as provided by the fit. 
The resulting $\Delta R/R$(t=2\,ps,$h\nu$)$_{\rm Drude}$ curve is shown in Fig.\,\ref{fig:TROS_fit}d). 
It shows that the contribution of the Drude term to the $\Delta R/R$ in the near-IR/visible spectral range is a nearly-flat positive signal. 
Hence, we argue that the large, positive signal in the near-IR spectral region is due to the modification of a Lorentz oscillator (namely, at 12300\,cm$^{-1}$ or 1.52\,eV) accounting for interband optical transitions. 
We can speculate on the origin of this optical transition by consulting the electronic band structure of (TaSe$_4$)$_2$I, as reported in \cite{Shi2021,Li2021}. 
Despite the fact that detailed calculations of the joint Density-of-States (JDOS) for (TaSe$_4$)$_2$I are not presently available in the literature, it is clear that optical transitions at photon energies of the order of or larger than $\approx$1.5\,eV link the quasi-continuum of nearly-flat bands found below the Fermi level (starting at an energy 0.5-0.6\,eV below E$_{\rm F}$) to another quasi-continuum of nearly-flat bands located above the Fermi level (starting at an energy 0.6-0.7\,eV above E$_{\rm F}$). 
This particular band structure is the reason why, for both parallel and perpendicular component, there is a large number of optical oscillators at rather high photon energy (overall from 10800\,cm$^{-1}$ to 28200\,cm$^{-1}$, that is, with photon energy in excess of $\approx$1.3\,eV), that links these two sets of bands. 
Other two oscillators at 4500\,cm$^{-1}$ and 7000\,cm$^{-1}$, as determined from the analysis of the perpendicular component of the optical properties, are likely linked to optical transitions from the set of bands below E$_{\rm F}$ to the states near E$_{\rm F}$ and from the states at E$_{\rm F}$ to the set of bands above E$_{\rm F}$. 
Because of the photon-energy range accessible in our time-resolved experiment, we can infer little about the contribution to the $\Delta R/R$ from these two low-energy optical transitions. 
However, from the analysis of the $\Delta R/R$ along the perpendicular direction, we can learn interesting details. 
For what was said above, it is not surprising that the only optical oscillator showing a modification of its central wave-number is the one with the lowest central wave-number ($\omega_0$, equal to 10800\,cm$^{-1}$ or 1.34\,eV) among the high-energy ones. 
Indeed, this energy acts as an effective energy-gap between the two sets of bands. Its transient modifications signify a sort of band-gap renormalization. 
After a first ultrafast increase, $\omega_0$ falls below the initial value, signalling a reduction of the effective band gap (this effect is responsible for the negative $\Delta R/R$ signal observed just after t$_0$), and then overshoots the initial value on a few-picoseconds timescale. The fact that the plasma frequencies of three oscillators (10800\,cm$^{-1}$, 15500\,cm$^{-1}$, 17000\,cm$^{-1}$) all vary indicates that photoexcitation triggers charge-transfer processes among different bands.

Finally, we comment on the timescales of the recorded modifications of the parameters, reported in Fig.\,\ref{fig:parameters}, starting from the parallel-direction terms. 
All the traces have been fitted with one or two exponential decays convoluted with a Gaussian function accounting for the time resolution, as also shown in Fig.\,\ref{fig:parameters}. 
After a slow buildup, the modification of the plasma frequency of the Drude term recovers on a timescale of (22$\pm$7)\,ps (the large uncertainty is due to the pump-probe delay range available). 
The scattering rate also shows a slow-buildup, and recovers its equilibrium value on a slightly shorter timescale, (12$\pm$7)\,ps.
The dynamics of the parameters of the 12300\,cm$^{-1}$ interband oscillator are different, and display a faster relaxation followed by a plateau-like decay. The central-frequency relaxes in (900$\pm$200)\,fs (with a second, slower decay of (50$\pm$10)\,ps). 
The plasma frequency shows a plateau-like behavior only, while the linewidth displays a complex dynamics. 
For what concerns the time-evolution of the perpendicular direction terms, we already described qualitatively the evolution of the central wave-number $\omega_0$ of the oscillator at 10800\,cm$^{-1}$. 
Quantitatively, after a first transient, faster than the time resolution and likely associated to the imperfections in the correction of the probe pulse chirp, $\omega_0$ evolves with an exponential time-constant of (3.75$\pm$0.15)\,ps. 
The plasma frequency of the oscillators at 10800\,cm$^{-1}$ and 17000\,cm$^{-1}$ decays with a (1.3$\pm$0.15)\,ps and (1.8$\pm$0.15)\,ps respectively, then they converge to a plateau. 
The plasma frequency of the oscillator at 15500\,cm$^{-1}$ instead converges to a value smaller than at equilibrium in $\approx$40\,ps. Finally, the linewidth of the oscillator at 17000\,cm$^{-1}$ shows a time-constant of (5.3$\pm$0.15)\,ps.

\subsection{X-ray Excitation}
\begin{figure*}[t]
\includegraphics[width=0.92\textwidth]{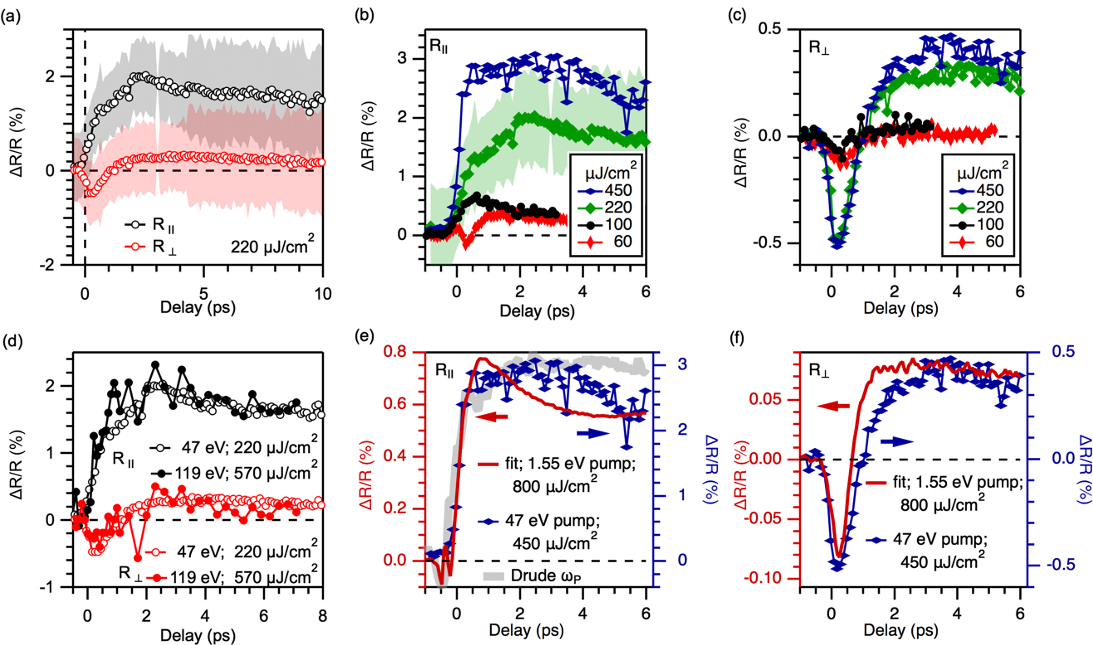}
\caption{\footnotesize FEL-induced optical reflectivity changes.
(a) Reflectivity changes for R$_{\parallel}$ (black open circles) and R$_{\perp}$ (red open circles) at RT measured for an absorbed fluence of 220\,$\mu$J/cm$^2$ and an excitation energy of 47\,eV. The shadowed areas mark the standard deviations of the signal. The FEL flux was estimated from the nominal transmission of the beamline.
(b) Fluence series measured for $\Delta$R$_{\parallel}$/R$_{\parallel}$ for a pump photon energy of 47\,eV.
(c) Fluence series measured for $\Delta$R$_{\perp}$/R$_{\perp}$ for a pump photon energy of 47\,eV.
(d) Comparison of the transient changes of R$_{\parallel}$ and  R$_{\perp}$, respectively, for two different excitation photon energies. 
(e) and (f) show comparisons of the FEL induced reflectivity changes (blue markers) with the changes expected for a probe photon energy of 1.55\,eV after excitation with 1.55\,eV (red curve) in parallel and perpendicular direction, respectively. 
The red traces for the optical excitation data are extracted from the maps shown in Fig.\,\ref{fig:TROS_fit}b) and c). 
As a guide to the eye, we included the temporal evolution of the Drude plasma frequency (grey shaded curve) to panel e).}
\label{fig:FEL_pp}
\end{figure*}
After having presented the results obtained for an optical pumping, which can only excite the valence band, we compare them with the non-equilibrium data obtained for FEL excitation, which, in addition to the valence band excitation, enables an initial excitation mechanism involving a core hole state.
These experiments were performed with a single-color probe beam at 800\,nm, as it was schematically shown in Fig.\,\ref{fig:setups}d). 
Figure\,\ref{fig:FEL_pp}a) shows the differential reflectivity data obtained at room temperature along the direction of chain formation (metallic direction) R$_{\parallel}$ (black open circles) and perpendicular to this direction (insulator direction) R$_{\perp}$ (red open circles).
The shadowed areas mark the standard deviation of the signal for independent measurements.
For an absorbed X-ray fluence of 220\,$\mu$J/cm$^2$ R$_{\parallel}$ shows an increase in reflectivity of about 2\,\%.
Note that the FEL flux was estimated from the nominal transmission of the beamline.
The maximum change is reached within the first 2\,ps after excitation and is followed by a slow decay.
On the contrary for R$_{\perp}$ we observe a rapid reduction of the relflectiviy at 800\,nm that reaches its minimal value around 160\,fs after excitation and a sign change after ca. 1.5\,ps. The maximum increase of reflectivity is reached after ca. 3\,ps for the highest fluences and then decays towards equilibrium on the timescale of several ps.
We can compare the general line shape of the transients to former works discussing X-ray induced optical reflectivity for different types of materials.\cite{Durbin2012}
It can be noted that the division of the material response in a fast component (<\,1\,ps) and a second, slower component (several ps) is typically observed in insulating or semiconducting materials. 
Eckert {\it et al.}\cite{Eckert2015} discussed the changes in the refractive index in the case of X-ray excited Si$_3$N$_4$ and concluded that the fast response is given by the modification of both real and imaginary part of the refractive index, whereas the ps-behavior is dominated by real-index changes.
They furthermore observed that the sensitivity for probing the fast or slow component depends strongly on the incident angle of the probe beam. 
In contrast, for metallic materials, such as gold, no long timescale component was observed.\cite{Krupin2012}

In the case of R$_{\perp}$ we were measuring along a direction of the (TaSe$_4$)$_2$I crystal band structure where it has an energy-gap of ca. 1.3\,eV\cite{Shi2021}, which is lower than the probe photon energy of 1.55\,eV.
A similar outcome as in our case, was seen on GaAs by Gahl {\it et al.}\cite{Gahl2008} and Krupin {\it et al.}\cite{Krupin2012}.
In the semiconducting material they observed as well an initial drop in the reflectivity followed by a zero crossing, and a similar fluence dependence, as we measure for (TaSe$_4$)$_2$I along the $a$-axis, showing a simultaneous decrease of the dip intensity and the overall decrease in reflectivity on the ps scale with decreasing fluence.
Gahl {\it et al.} fitted the timescale for the initial drop in reflectivity to $\tau=(160\pm44)$\,fs, which is similar to our observation in the case of (TaSe$_4$)$_2$I.
The interpretation given for the fast drop of the reflectivity in GaAs after X-ray excitation is assigning it as a consequence of a band gap narrowing effect.\cite{Durbin2012}
In the case of band gap narrowing the number of valence band electrons that can be excited to the conduction band by the probe pulse is enhanced. 
In this way the absorption increases and the reflectivity decreases.
After the initial population of excited carriers in the conduction band, thermalization of the hot electrons sets in, leading to a filling of the conduction band bottom.
The band filling will block further absorption and hence the reflectivity will increase again.
The characteristic transient of an initial fast drop followed by a net increase of the reflectivity after few ps is hence explained by the simultaneous contribution of two different effects: band gap narrowing and band filling.
Finally electron-hole recombination leads to the decrease in reflectivity towards the equilibrium value on longer timescales.
As Durbin \cite{Durbin2012} points out, there are also other possible explanations for the initial dip in the reflectivity besides band gap narrowing, which is a pure effect of the valence/conduction band structure and hence an argument that comes directly from a comparison to optical spectroscopy, but does ignore the presence of core holes in case of X-ray excitation.
In the case of the semiconductor GaAs he describes the significant hole-density induced by X-ray excitation in analogy to semiconductor doping, which would also induce band gap narrowing effects. 
Furthermore, he mentions effects caused by an interference of the effects that both X-ray and optical pulse have on the modification if the density of excited conduction electrons.
Note that the very same shape of the transient after excitation was also observed in the large band-gap insulator Si$_3$N$_4$, where the net positive reflectivity on the ps-scale was explained by the formation of metastable electron-hole bound states inside the band gap.\cite{Casolari2014}
For a deeper analysis of our data one would need to model the modification of the dielectric function due to the X-ray excitation, as it was done by Tkaschenko {\it et al.}\cite{Tkachenko2016} for the example of some semiconducting materials using the XTANT code developed at DESY, which is an ongoing field of research.

Also the shape of the transients measured for R$_{\parallel}$, which should reflect the metallic character of the sample along this directions, resembles best the expected curves for X-ray excited metals, which was described by Durbin \cite{Durbin2012} for gold. 
In the case of gold an abrupt increase of the reflectivity is measured and a subsequent decay towards equilibrium within few picoseconds.\cite{Krupin2012,Durbin2012}
The decay times we are observing for the case of (TaSe$_4$)$_2$I are significantly longer, but nontheless show a strong fluence dependence. 
The influence of the X-ray fluence on the optical reflectivity changes in gold was not discussed in the former mentioned works and can hence not be used for comparison.
As Durbin states, the reason for the observed X-ray induced reflectivity changes in the case of gold is still under discussion. 
In the case of gold he is discussing inelastic scattering processes of the photoelectrons, which are promoting d-electrons into the unoccupied parts of the s-p bands, and arguing that the Drude reflectivity is not changing by increasing the number of conduction electrons.

The measurements shown so far were obtained using an FEL wavelength of 26\,nm, which corresponds to a photon energy of $\approx 47$\,eV.
Referring to ultraviolet photoemission studies, with a photon energy of 47\,eV electrons from the Ta 4f and 5p states will be emitted due to direct photoemission, whereas using higher photon energies also the I 4d and Se 3d states can be addressed.\cite{Ohtake1986} 
Furthermore, Othake {\it et al.}\cite{Ohtake1986} observed that using a photon energy of 48\,eV results in an enhanced partial cross-section for the detection of Ta 5d valence electrons in the angle-integrated UPS spectra, which would give an impact to the optical reflectivity.
Figure\,\ref{fig:FEL_pp}d shows the comparison of photoexcitation with photon energies of 47\,eV and 119\,eV and comparable amount of photons per pulse. 
Both transients are looking very similar.
Even though we can address further core levels with the higher photon energy and probably excite less Ta 5d electrons, we do not observe a significant impact on the conduction and valence band structure. 

The overall shape of the reflectivity changes observed for X-ray excitation resembles our observations from optical pump-probe spectroscopy. 
Figures\,\ref{fig:FEL_pp}e) and f) show a direct comparison of the traces of the reflectivity changes along the two directions for excitation with 47\,eV photons and an absorbed fluence of 450\,$\mu$J/cm$^2$ (blue markers) and the extrapolated behavior observed for an excitation energy of 1.55\,eV photon energy and an absorbed fluence of 800\,$\mu$J/cm$^2$ (red traces) for the same probe photon energy. 
The traces for the 1.55\,eV pump case are extracted from the maps shown in Figs.\,\ref{fig:TROS_fit}b),c).
The amplitude of the recorded $\Delta$R/R signal is much higher in the case of the X-ray pump energy (cf. right axes) than for near-IR excitation. 
We observe maximal changes in the reflectivity of 3\,\% in $\parallel$ direction and $\pm$0.5\,\% in perpendicular direction for FEL excitation, whereas they are $\leq$1\,\% and $\leq$0.01\,\% for near-IR excitation.
Since in both cases the probe-beam was hitting the sample at near-normal incidence, we ascribe these variations to the difference in the pump wavelength rather than to experimental geometry effects. 
However, the rising edge of the initial change in reflectivity is identical for both types of excitations. 
This is reflected along both measurement directions.
Since the core-hole lifetimes are typically in the order of only few fs, the time-resolution in our experiments does not allow for distinguishing if the pump pulse directly access the conduction band, or secondary processes after a core-hole generation modify the carrier density in the conduction band.
Concentrating on the perpendicular direction, we observe a slightly earlier appearing compensation point of the band-gap narrowing and band-filling processes in the case of optical excitation. 
However, the traces along the perpendicular direction are very similar in both experiments.
In contrast for the parallel direction we see different decay dynamics of the enhanced reflectivity due to the excitation.
In the case of X-ray excitation with increasing fluence we observe a second, slower timescale for reaching the maximal change in reflectivity (cf. green curve in Fig.\,\ref{fig:FEL_pp}b)), or a plateau-like behavior (cf. blue curve in Fig.\,\ref{fig:FEL_pp}b)) that decays only after few picoseconds.
On the contrary for valence-band excitation with 1.55\,eV photons we reach the maximal change in reflectivity in less than 1\,ps and observe an intermediately initiating decay until it reaches ca. 75\,\% of the highest change and saturates.
This behavior initially resembles the temporal evolution of the plasma frequency of the Drude term, that was extracted from the optical spectroscopy experiments.
As demonstrated by the grey shaded curve in Fig.\,\ref{fig:FEL_pp}e), the Drude plasma frequency perfectly describes the dynamics of the first 4\,ps after FEL excitation.
The reflectivity is then decreasing, which means that some Lorentz-oscillators must be affected as well as a consequence of the initial core-hole excitation.
In the optical spectroscopy data, the Drude term gives a broadband offset to the reflectivity, whereas the fast dynamics is dominated by the oscillator at 12300\,cm$^{-1}$, to which we are close to resonance when using a pump photon energy of 1.55\,eV.
This finding suggests that the core-level excitation affects mainly the free carriers (metallic conduction electrons), that subsequently decay in the conventional way through electron-phonon scattering processes, while the excitation of interband optical transitions is somehow less effective in a metallic environment by secondary processes after a core-level excitation.

\begin{figure}[t]
    \centering
    \includegraphics{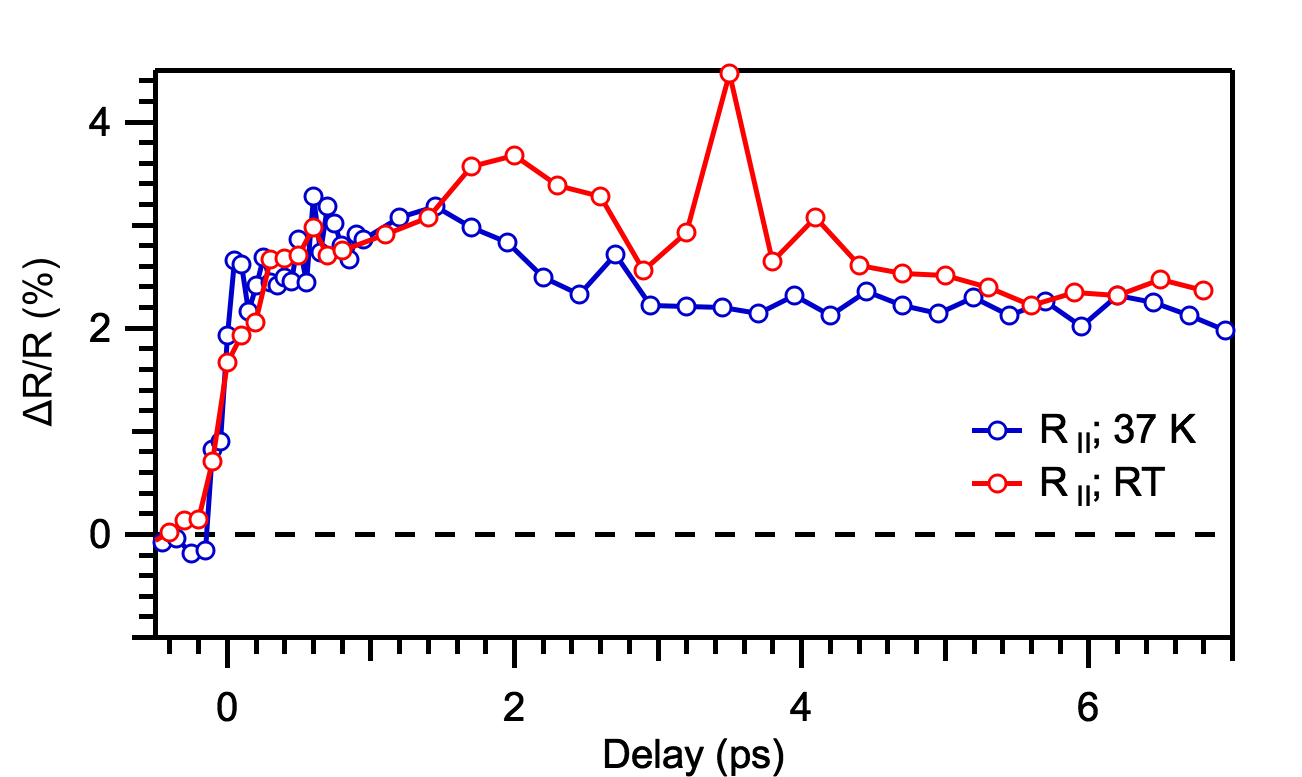}
    \caption{\footnotesize Comparison of the transient changes of R$_{\parallel}$ above and below the critical temperature for charge density formation for an excitation energy of 119\,eV.}
    \label{fig:comp_Temp}
\end{figure}
We were furthermore interested in investigating if the core-level excitation can couple to the charge-density wave, which is present at sample temperatures below 263\,K.
Optical excitation of the valence band structure leads to the excitation of the amplitude mode (the phonon mode coupled to the structural distortion induced by the CDW), and a faster initial relaxation of the reflectivity change of R$_{\parallel}$, as discussed in Sec.\,\ref{sec:optical pump-probe}.
Both effects are not present when exciting with an FEL pulse, as can be seen in the direct comparison of the $\Delta R/R$ transients obtained at RT and at 37\,K shown in Fig.\,\ref{fig:comp_Temp}.


\section{Conclusions}
The comparison of optical reflectivy changes induced by optical and X-ray excitations for the quasi one-dimensional material (TaSe$_4$)$_2$I along its $a$- and $c$-axes showed that the qualitative response along the insulator direction does not depend on the excitation pathway, whereas along the chain direction, the decay mechanisms of the initial reflectivity change are different. 
Furthermore, we are not able to detect fingerprints of the charge-density wave phase in the $\Delta R/R$ signal when using core-level excitations.
At the present stage, we cannot conclude whether this is due to the X-rays destroying the CDW or due to their excitation pathway not coupling directly to the CDW. 
Thus, we think that our results can be of high interest for theoretical studies on X-ray induced optical reflectivity changes, which is a rather new field of investigation.\\
An interesting future experiment to explore in more detail the variations of the dielectric function after X-ray excitation would be to set up an X-ray pump/ optical broadband-probe experiment. 
Data gained from such an experiment would allow for performing an analysis as discussed in Sec.\,\ref{sec:optical pump-probe} for optical pump-probe spectroscopy and hence to identify the modified oscillators after excitation.
This would provide insight in the reason for the different de-excitation dynamics we are observing for valence- and core-level excitation along the $c$-axis in (TaSe$_4$)$_2$I.
Finally, the study of the influence of photoexcitation via different pathways on the non-equilibrium properties of complex materials is an emerging field that will provide new insights about the functional properties of metallic and insulating phases and their possible control.

\section*{Author Contributions}
F. C. designed and performed the time-resolved table top optical spectroscopy experiments. M. T., G. C., M. De C. and F. C. contributed to the analysis of the time-resolved purely optical spectroscopy data. 
W. B. and F. C. had the experimental lead of the experiments performed at FERMI to study FEL-induced optical reflectivity changes.
H. B. grew the samples.
G. K. took care of the optical setup at the TIMER endstation.
W. B., M. T, M. de C., D. P., D. S., A. G., M. P., D. De A., D. F., E. P., S. P. C., R. M., L. F., F. B. and F. C. performed the experiments on FEL-induced optical reflectivity changes.
W. B. analyzed the data gained in the FEL experiments.
F. P. contributed to the scientific discussion.
W. B. and F. C. wrote the original paper draft, which was proofread by all co-authors.

\section*{Conflicts of interest}
There are no conflicts to declare.

\section*{Acknowledgements}
We acknowledge the contribution of Simone Peli to the original beam-time proposal allowing us to perform the experiments on X-ray induced optical reflectivity measurements.
Furthermore we thank Alberto Crepaldi for the fruitful discussions and his support in setting up the collaboration with the sample growers.\\
E. Paltanin acknowledges the contribution of the European Union's Horizon 2020 research and innovation program under the Marie Sklodowska-Curie grant agreement No 860553.

\bibliographystyle{apsrev4-1}
\bibliography{TaSebib}
\end{document}